\documentclass[floatfix,superscriptaddress,a4paper,
               showpacs,showkeys,nofootinbib,preprint]{revtex4}
\usepackage{epsfig}
\usepackage{latexsym}
\usepackage{xspace}
\usepackage{hyperref}
\usepackage[latin2]{inputenc}
\usepackage{indentfirst}
\usepackage{enumerate}
\usepackage{color}

\usepackage{amsmath}
\usepackage{amssymb}
\usepackage[english]{babel}
\usepackage{url}
\graphicspath{{./graphics/}}

\newcommand{\eq}[1]{\begin{align} #1 \end{align}}

\begin{document}

\title{Nucleon matter equation of state, particle number fluctuations,\\ and shear viscosity within UrQMD box calculations}

\author{A. Motornenko}
\affiliation{Frankfurt Institute for Advanced Studies, Giersch Science Center, Frankfurt am Main, Germany}
\affiliation{Institut f\"ur Theoretische Physik,
Goethe Universit\"at Frankfurt, Frankfurt am Main, Germany}

\author{L. Bravina}
\affiliation{Department of Physics, University of Oslo, Oslo, Norway}

\author{M. I. Gorenstein}
\affiliation{Frankfurt Institute for Advanced Studies, Giersch Science Center, Frankfurt am Main, Germany}
\affiliation{Bogolyubov Institute for Theoretical Physics, Kiev, Ukraine}

\author{A. G. Magner}
\affiliation{Institute for Nuclear Research NASU, Kiev, Ukraine}

\author{E. Zabrodin}
\affiliation{Department of Physics, University of Oslo, Oslo, Norway}
\affiliation{Skobeltsyn Institute of Nuclear Physics, Moscow State
University, Moscow, Russia}
\affiliation{National Research Nuclear University "MEPhI" (Moscow
Engineering Physics Institute), Moscow, Russia}

\begin{abstract}
Properties of equilibrated nucleon system are studied within
the Ultra-relativistic Quantum Molecular Dynamics (UrQMD) transport model.
The UrQMD calculations are done within a finite box with periodic boundary conditions.
The system achieves  thermal equilibrium due to nucleon-nucleon elastic scattering.
For the UrQMD equilibrium state, nucleon energy spectra, equation of state, particle number
fluctuations, and shear viscosity $\eta$ are calculated. The UrQMD results are
compared with both, statistical mechanics and Chapman-Enskog kinetic theory,  for a classical system
of nucleons with hard-core repulsion.

\end{abstract}



\maketitle
\centerline\today

\section{Introduction}
Relativistic transport theory is a well established
approach for description of multi-particle hadronic dynamics.
Most transport models are based on a microscopic description of the non-equilibrium hadronic
stage of nucleus-nucleus (A+A) collisions.  In what follows we will employ the  Ultra-relativistic Quantum
Molecular Dynamics (UrQMD) model \cite{Bass:1998ca,Bleicher:1999xi}.

The final stages of A+A collisions can be effectively well described within statistical models.
These models assume that the
final particles are emitted from an equilibrated system that can be described by just
a few parameters: the system volume $V$, chemical potential(s) $\mu$, and temperature $T$.
These thermodynamical parameters are usually found from fitting
measured multiplicities of different hadron species in A+A collisions.
In many cases, ideal  hadron-resonance gas models give a satisfactory description.
To consider the effects of inter-particle repulsion,
the excluded volume (EV) model was suggested a long
time ago \cite{Rischke:1991ke,Cleymans:1992jz,Yen:1997rv,Gorenstein:1999ce}.
In this model, the hard-core repulsion between particles is considered by using
the van der Waals excluded volume procedure.

The present paper aims to provide a better understanding of the correspondence between
UrQMD and statistical models in A+A collisions.
This purpose can be achieved with consideration of the both models at the same
conditions, i.e., at the thermodynamical equilibrium. Studies of the equilibrated multi-component
hadronic systems were performed by using the UrQMD model within a box with periodic
boundary conditions \cite{Belkacem:1998gy}, and for dynamical systems created in
the central cell of  A+A collisions \cite{Bravina:1998pi,Bravina:1999dh,Bravina:2000iw}.
However, a proper comparison of the UrQMD results with
predictions of the statistical model is rather problematic, as in the
UrQMD model the conditions of detailed balance are violated by multi-body decays
of string degrees of freedom and some resonances.

The present work considers system of nucleons at low energies (temperatures $T\le 50$ MeV)
in a box with periodic boundary conditions.
The system reaches thermal equilibrium via nucleon-nucleon elastic scatterings.
Since the average energy per nucleon is small, only elastic collisions occur
and new degrees of freedom (e.g., $\pi$ and $\Delta$) are not excited and, thus,
the conditions of detailed balance are satisfied.
In addition, a nonrelativistic approximation can be adopted to the nucleon kinetic energy.
Note, that the standard version of the UrQMD model that was used in the present study
does not include mean-field interactions.

Particle collisions in the UrQMD model are described via the so-called ``black-disk'' mechanism.
This can be considered as an implementation of the particle finite size.
A  comparison between results from the UrQMD box calculations with statistical mechanics
of the EV model and kinetic model for hard spheres
will be presented for nucleon energy spectra, equation of state,
particle number fluctuations, and shear viscosity.

The main subject of our studies is a comparison of equilibrium properties of the nucleon system which are
calculated within the UrQMD transport model with the results of statistical mechanics and kinetics for the system of
hard spheres. Recently a detailed comparison
of different transport codes (including the UrQMD model) under controlled conditions
was presented in Refs.~\cite{Xu:2016lue} and \cite{Zhang:2017esm}.
These questions are, however, outside of the scope of the present paper.

The paper is organized as follows.
In Sec. \ref{sec:theory} the statistical EV model
is presented along with kinetic model
results for the shear viscosity $\eta$.
Section \ref{sec:results} shows the results of the UrQMD box calculations
and presents a comparison with those of the statistical EV model.
Section \ref{sec:viscosity} presents the UrQMD shear viscosity and its
comparison with Chapman-Enskog kinetic theory. A summary in
Sec. \ref{sec:summary} closes the article.

\section{Statistical equilibrium of hard-sphere nucleons}
\label{sec:theory}
Let us consider $N$ classical non-interacting particles in a
volume $V$ (both quantum and relativistic effects are neglected).
The statistical equilibrium corresponds to a homogeneous particle distribution
in the coordinate space, and Maxwell-Boltzmann distribution in the momentum space,
\eq{\label{M}
f(p)~\equiv~\frac{1}{4\pi V}\frac{dN}{p^2 dp}~= ~\frac{N}{V}~(2\pi mT)^{-3/2}~\exp\left(-~\frac{p^2}{2mT}\right)~,
}
where $p\equiv|{\bf p}|$ is the momentum and $m$ is the mass of a particle,
and $T$ is the system temperature. The function $f(p)$ is normalized as
\eq{\label{fnorm}
4\pi\int\limits_0^{\infty}p^2dp~f(p)~=~\frac{N}{V}~\equiv ~n~,
}
where $n$ is the particle number density. A particle's kinetic energy
equals to $\epsilon = p^2/(2m)$ and its average value is determined by the system temperature $T$,
\eq{\label{T}
\langle \epsilon \rangle~ \equiv~\frac{4\pi}{n}\int\limits_0^{\infty}p^2 dp~
\epsilon~f(p)~=~\frac{3}{2}\,T~.
}
The ideal gas pressure
is given by
\eq{\label{Pid}
P_{\rm id}~=~\frac{N\,T}{V}~\equiv ~n\,T~.
}

The statistical description of a gas with hard-core repulsive interactions between particles
can be done via the van der Waals EV procedure.
which yields the system's pressure:
\begin{equation}\label{PEV}
P_{\rm EV} ~= ~\frac{nT}{1-bn}~,
\end{equation}
where $b=16\pi r^3/3$ is a particle excluded volume (four times larger
than the own particle volume) and $r$ is
a particle hard-core radius\footnote{Note that
at typical temperatures for hadronic and nuclear physics the
quantum mechanical effects for nucleon-nucleon interactions
neglected in the present paper can be important
(see recent  Ref. \cite{Vovchenko:2017drx}).}.
Equation (\ref{PEV}) is valid at the condition $bn\ll 1$,
for higher densities the Carnahan-Starling \cite{CarnahanStarling} model
for the gas of hard spheres can be applied.
Note that the hard-core repulsion
does not modify Eqs.~(\ref{M}-\ref{T}).

Particle number fluctuations in the grand canonical ensemble (GCE) are sensitive
to the interaction between particles. The scaled variance $\omega[N]$ for the EV
equation of state (\ref{PEV}) can be calculated as \cite{Gorenstein:2007ep,Vovchenko:2015pya}:
\begin{equation}
\label{eq:omega}
\omega[N]~\equiv~\frac{\langle N^2\rangle -
\langle N\rangle^2}{\langle N\rangle}~=~(1-b\,n)^2~,
\end{equation}
where $\langle ... \rangle \equiv \sum_N ... W(N)$, and $W(N)$ is the
particle number probability distribution.
The expression (\ref{eq:omega}) shows a suppression of particle number
fluctuations with increasing density $n$. This property is in contrast
to the ideal gas where $\omega[N]=1$, and is independent of $n$.

Another quantity that is related to the particle interaction is
the shear viscosity $\eta$.
The shear viscosity $\eta$ describes the momentum transfer
due to the particle thermal motion,
and depends on the particle elastic scattering.
For hard-sphere particles $\eta$ was estimated by Maxwell
as
\eq{\label{eta}
\eta\sim n\,m\, l\,v_{\rm th}\sim \frac{\sqrt{mT}}{d^2}~,
}
where $l \sim (nd^2)^{-1}$ is the mean free path of a particle between
two successive collisions, $v_{\rm th}\sim\sqrt{T/m}$ is a particle
thermal velocity, $d=2r$ is the particle's hard-core diameter. An accurate
expression of $\eta$ in the system of hard-sphere particles
was obtained by Chapman and Enskog
\cite{Chapman:1951} in the so-called frequent
collision (FC) regime, i.e., when a system size $L$ is much larger than the
mean free path, $L\gg l$,
\begin{equation}
\eta^{}_{\rm FC} ~=~ \frac{5}{16} \frac{\sqrt{\pi m T}}{\sigma_{\rm int}}~.
\label{eq:CE-sigma}
\end{equation}
Here $\sigma_{\rm int}=\pi d^2$ is the hard-sphere elastic cross section.
Note that this expression for the viscosity $\eta_{\rm FC}$ is independent of the particle density $n$.

The shear viscosity in the ideal gas limit $nd^3\rightarrow 0$
is not well defined. In this limit, one has $l\rightarrow \infty$. If the requirement $L\gg l$ remains
valid, one finds $\eta_{\rm FC}\rightarrow \infty$ from Eq.~(\ref{eq:CE-sigma}).
However, at any fixed value of $L$, the so-called rare collision (RC) regime
with $l\gg L$ takes place
which leads to $\eta_{\rm RC}\rightarrow 0$ in this limit
(see Ref.~\cite{Magner:2017ewj}).

\section{UrQMD box simulations of equilibrated nucleon matter}
\label{sec:results}

\subsection{General UrQMD ingredients}
\label{sec:UrQMD}

In the present work we use the standard implementation of the UrQMD model, i.e.,
mean-fields, two- and three-body particle potentials are turned off, and,
thus, dynamics of the system are described only through isotropic elastic scatterings. The particle
collision term includes Pauli-blocking suppression factor that is implemented to effectively treat Fermi-Dirac statistics.

An equilibrated nucleon gas is simulated with the UrQMD box calculations using
the periodic boundary conditions. An isotropic symmetric
system of $N$ nucleons is considered, i.e., $N/2$ neutrons and $N/2$ protons.
We neglect a difference between proton and neutron masses
in analytical estimates below and take $m=938$~MeV. The numerical values of the system parameters are presented in Table \ref{table1}.
\begin{table}[h!]
\centering
\setlength{\tabcolsep}{0.5em}
\begin{tabular}{c|c|c|c|c|c}
$N$ & $L$ (fm) & $n$ (fm$^{-3}$) & $\langle \epsilon \rangle$ (MeV)
& $\sigma_{\rm int}$ (mb) & $t_{\rm eq}$ (fm/c)\\
\hline \hline
400 & $9.3 - 34.2$ & $0.01 - 0.5$ & $15-75$ & $10 - 80$ & $ < 150$
\end{tabular}
\caption{Properties of a nucleon gas studied in the UrQMD box.}
\label{table1}
\end{table}
Note that in the UrQMD calculations we set the elastic cross section
to the same fixed value $\sigma_{\rm int}$ for both protons and neutrons, which is independent of
the collision energy of nucleon pairs. In calculations all inelastic
reactions are disabled. The values of $\sigma_{\rm int}$
from $10$~mb up to 80~mb are considered. For the hard-sphere cross
section  with $\sigma_{\rm int} =\pi d^2$ these values correspond to the
hard-core particle diameters from $d=0.56$~fm to $d=1.60$~fm, respectively.

The system is initialized as follows. In a cubic box with
the side $L=V^{1/3}$ we put $N=400$ nucleons (other larger values
of $N$ are considered only to study the particle number fluctuations)
with an isotropic distribution $f({\bf p})=f(p)$ and the step-like
shape $f(p)={\rm const}$ up to $p=p_{\rm max}$.
The value of $p_{\rm max}$ determines the value of the average energy
per nucleon, $\langle \epsilon \rangle=3\,p_{\rm max}^2/(10\,m)$.
The particle density $n$ is changed by considering different sizes
of the box side $L$ at the fixed $N$. Nucleon momentum spectra are not
changed with time at $t>t_{\rm eq}$. It has been found that
$t_{\rm eq} < 150$~fm/c for all considered combinations of
the system parameters.

\begin{figure}[h!]
\centering
\includegraphics[width=0.6\textwidth]{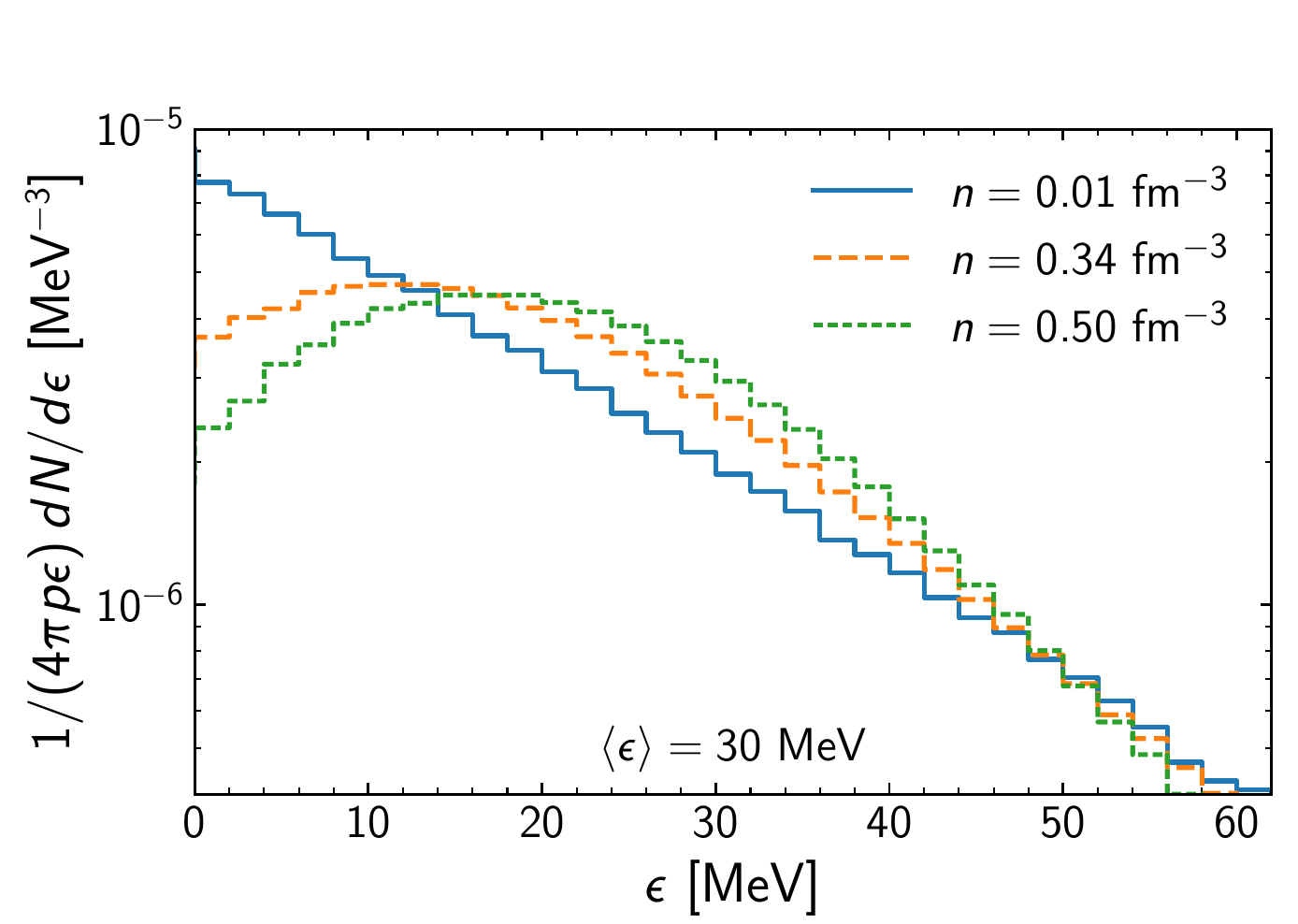}
    \caption{Energy spectra of nucleons from UrQMD box calculations for different
    nucleon densities. The mean nucleon energy is fixed to $\langle \epsilon \rangle=30$~MeV.}
\label{fig:en-spectr}
\end{figure}

In what follows we compare the UrQMD-equilibrium state at $t>t_{\rm eq}$ with
the statistical mechanics equilibrium of hard spheres.
As seen from Fig.~\ref{fig:en-spectr}, the nucleon energy spectra
have the Maxwell-Boltzmann shape (\ref{M}) at small density, whereas at large $n$ a deviation
from Eq.~(\ref{M}) takes place.
A modification of the energy spectra is a result of the Pauli-blocking mechanism
implemented in the
standard version of the UrQMD model. This implementation, however, does
not take into account genuine quantum nature of nucleons, and it results in the equilibrium nucleon spectra
which do not correspond to the Fermi distribution. Thus, in general case, the temperature parameter
can not be used to characterize the shape of nucleon equilibrium spectra in the box.
When the energy spectra extracted from the UrQMD calculations are different from Eq.~(\ref{M}),
we still use the mean particle energy $\langle \epsilon \rangle$ to define the effective
temperature $T$ according to Eq.~(\ref{T}), i.e., $3T/2$ will be considered as a universal measure of nucleon
kinetic energy at equilibrium.

\begin{figure}[h!]
\centering
\includegraphics[width=0.49\textwidth]{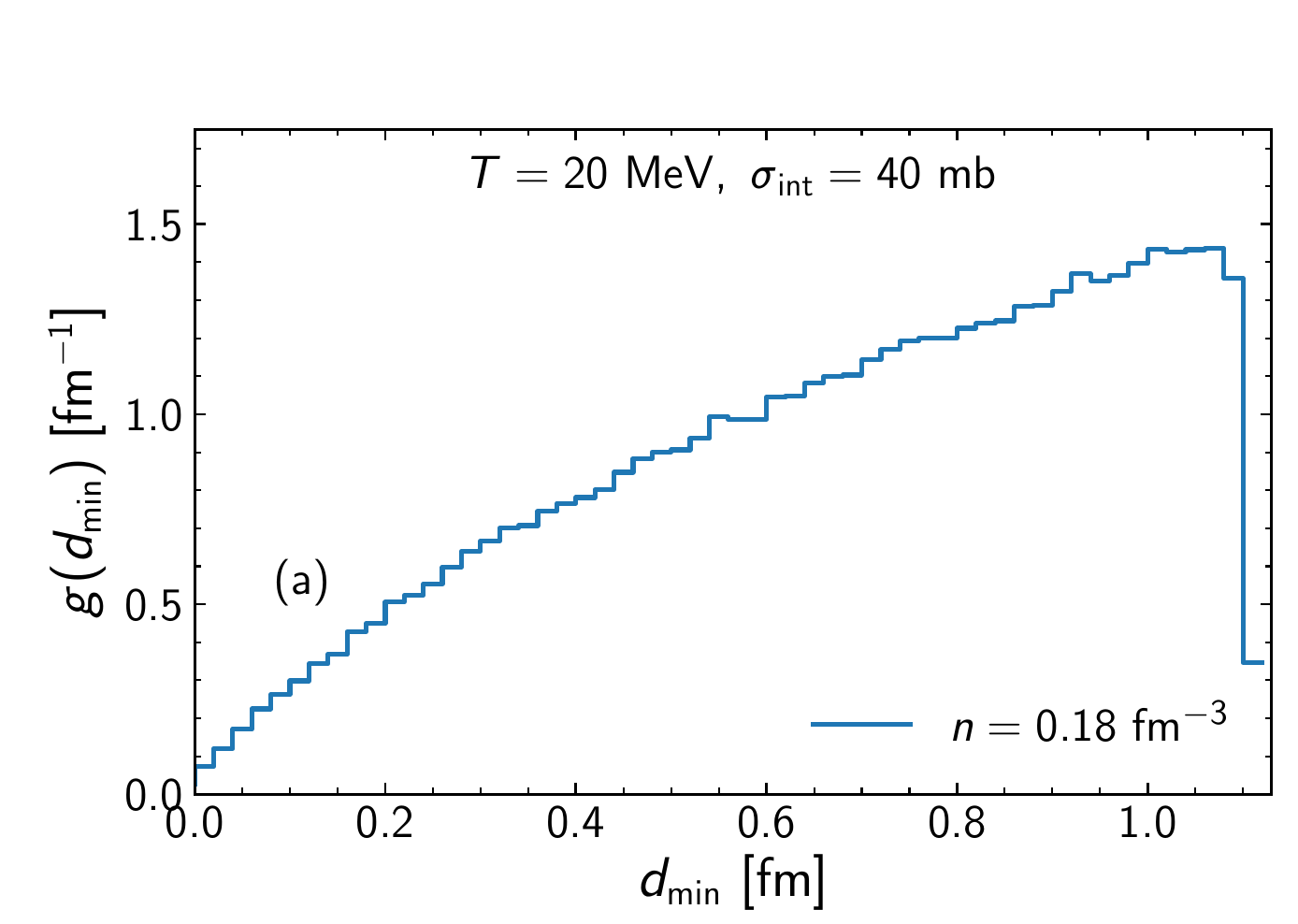}
\includegraphics[width=0.49\textwidth]{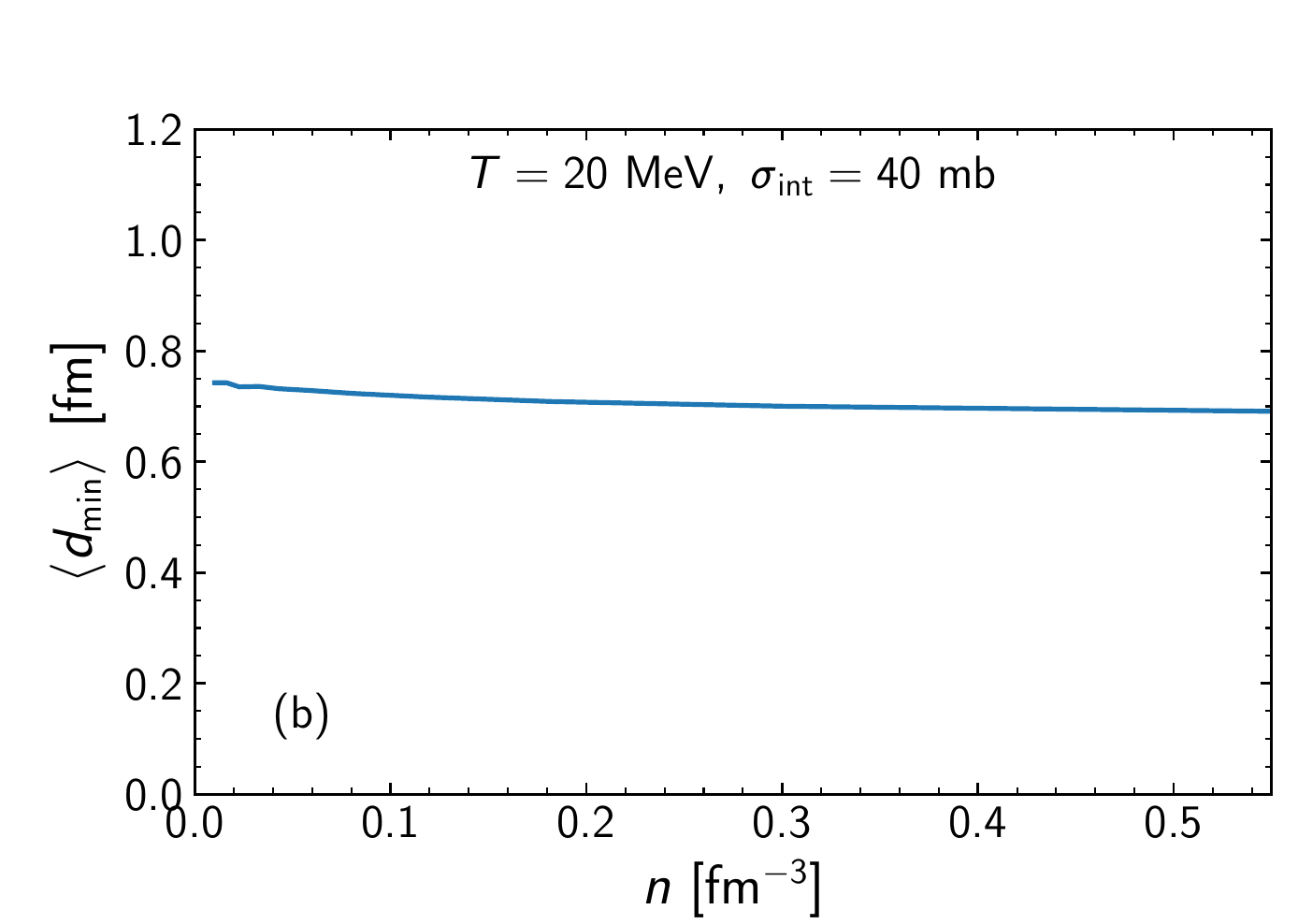}
    \caption{(a): The UrQMD results for the probability distribution $g(d_{\rm min})$
    of distances $d_{\rm min}$ between nucleons at the time of their interaction
    at the density $n=0.18$ fm$^{-3}$. (b): The average distance $\langle d_{\rm min}\rangle$
    as a function of the particle density $n$. The interaction cross section is
    fixed as $\sigma_{\rm int}=40$ mb, and temperature is $T=20$ MeV.}
\label{fig:mean-dist}
\end{figure}

In the UrQMD model the nucleon dynamics are treated
as the following. Particles move as free streaming between
successive collisions.
Each pair of particles is assumed to be free streaming until
the minimal distance $d_{\rm min}$ between them is reached.
Then, the value of $d_{\rm min}$
is compared with the interaction distance, $d_{\rm int} = \sqrt{\sigma_{\rm int}/\pi}$.
If  $d_{\rm min}\le d_{\rm int}$,  the elastic scattering
takes place. 
This is different from the elastic scattering of hard spheres where
the distance between their centers can not be smaller than  the diameter of the spheres.
The probability distribution $g(d_{\rm min})$ of $d_{\rm min}$ at the equilibrium stage $t>t_{\rm eq}$
is presented in Fig.~\ref{fig:mean-dist}~(a) at a given particle density,
while the mean value of $\langle d_{\rm min}\rangle  $ is plotted versus the particle density $n$
in Fig.~\ref{fig:mean-dist}~(b). As $\langle d_{\rm min}\rangle  $ shows approximately a constant
value independent on the nucleon number density,
one may expect that UrQMD nucleons would behave according to Eqs.~(\ref{PEV}) and (\ref{eq:omega})
similar to the system of spheres with
a diameter $d$ equal to
$\langle d_{\rm min}\rangle$.  However, in contrast to these expectations, both the pressure $P$ and scaled variance $\omega[N]$
in the UrQMD box calculations demonstrate the pure ideal gas behavior.

\subsection{Pressure calculation}
\begin{figure}[h!]
\centering
\includegraphics[width=0.6\textwidth]{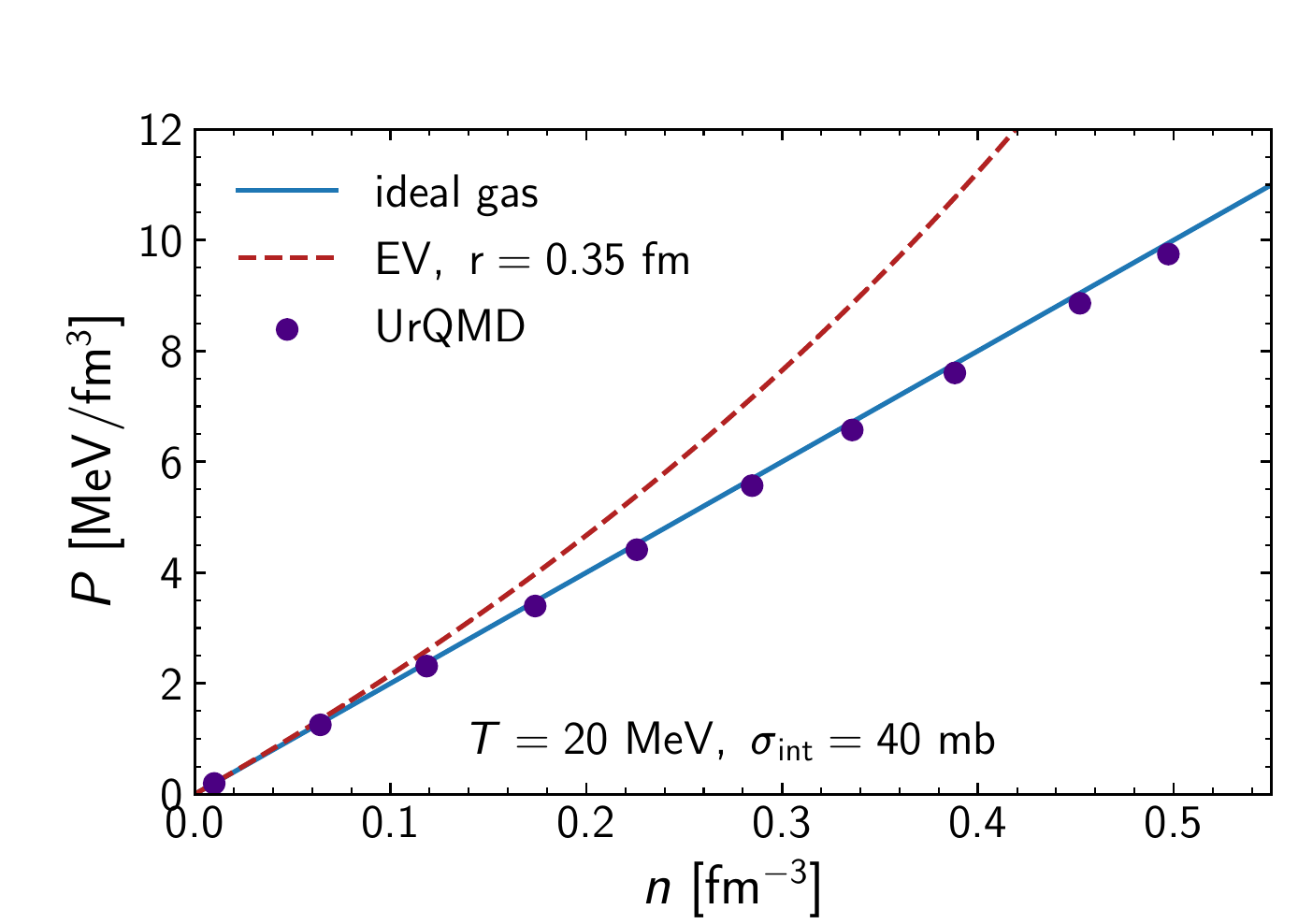}
    \caption{Pressure $P$ of the nucleon gas from UrQMD calculations is
    shown by circles as a function of density $n$. Solid and dashed lines present
    the ideal gas (\ref{Pid}) and the EV (\ref{PEV}) behavior, respectively.}
    \label{fig:pressure}
\end{figure}

The pressure $P$ in the UrQMD-equilibrium state can be calculated as a trace
of spatial components of the stress-energy tensor $T^{ij}$ \cite{Belkacem:1998gy}:
\begin{equation}\label{box-pressure}
P~=~\frac{1}{3}\,\sum\limits_{i=1}^{3}T^{ii}~=~\frac{1}{3}\,
\sum\limits_{i=1}^{3}\langle p_k^i v_k^i\rangle\,\equiv~\frac{1}{V}\,\frac{1}{\rm N_{ens}}
\sum_{ h=1}^{\rm N_{ens}}\sum_{k=1}^N \sum_{i=1}^3(p^i_k v^i_k)_h~,
\end{equation}
where $p^i_k$ and $v^i_k$ denote the $i$th projection of $k$th particle
momentum and velocity, respectively, and
$h$ enumerates the microscopic states within the ensemble of ${\rm N_{ens}}$ states.
Figure \ref{fig:pressure} shows the UrQMD-equilibrium  pressure (\ref{box-pressure})
as a function of $n$. The parameter $T$ at all $n$ is fixed according to Eq.~(\ref{T}).
The UrQMD  pressure (\ref{box-pressure}) coincides with the ideal gas pressure
(\ref{Pid}). Therefore, no EV effects
are seen\footnote{In Refs.~\cite{Nara:2016hbg,Nara:2016phs,Nara:2017qcg}
the constraints on the two-body scattering were implemented into the molecular dynamics code
to model the non-ideal equation of state. These constraints are however absent in the standard UrQMD
code used in the present paper.}.
The ideal gas behavior of the UrQMD pressure remains valid at
large densities $n$ with no manifestations of the EV effects due to a nonzero value of $\langle d_{\rm min}\rangle$.
The EV pressure function (\ref{PEV}) with $d=0.35$~fm is shown in
Fig.~\ref{fig:pressure} for a comparison.

Another reason that could change the UrQMD pressure is the behavior
of nucleon energy spectra at large $n$  seen in Fig.~\ref{fig:en-spectr}.
As it was mentioned earlier, the energy spectra of nucleons at
large $n$ are rather different from the Maxwell-Boltzmann distribution (\ref{M}).
Equation (\ref{Pid}) is nevertheless valid for the UrQMD pressure.
This  happens because of our agreement to use Eq.~(\ref{T}).
Indeed, calculating the UrQMD pressure by Eq.~(\ref{box-pressure}), one obtains
\eq{\label{press}
P~=~\frac{1}{3V} \sum_{k=1}^{N} \frac{p_k^2}{m}~=~\frac{2N}{3V} \langle \epsilon \rangle ~=~n\,T~,
}
where Eq.~(\ref{T}) is used at the final step.

\subsection{Particle number fluctuations }
\begin{figure}[h]
\centering
\includegraphics[width=0.6\textwidth]{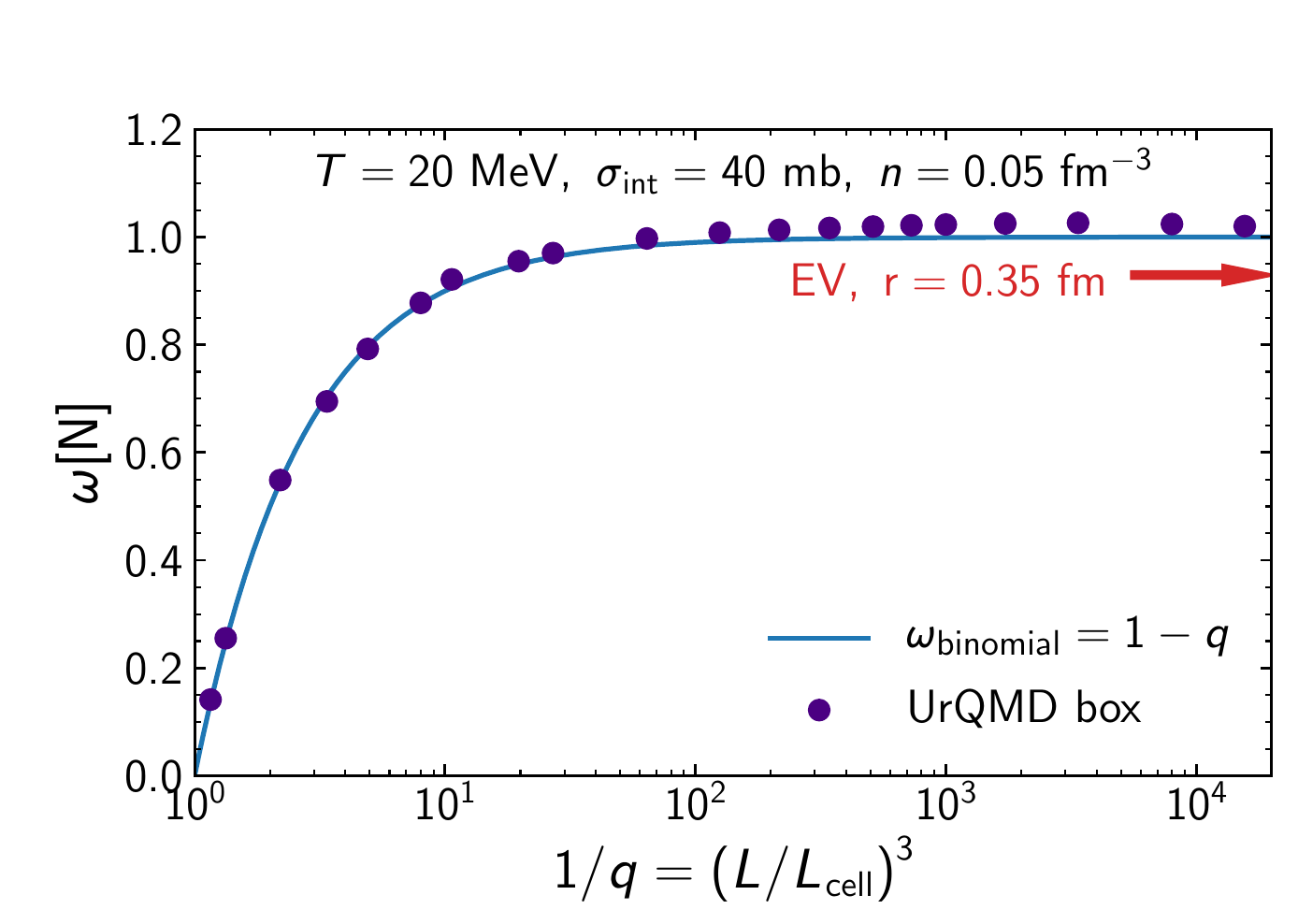}
    \caption{The scaled variances $\omega[N]$ of  particle number
    fluctuations inside the cell as a function of $1/q=(L/L_{\rm cell})^3$.
    A solid line is the
    binomial distribution (\ref{eq:binom}),
    an arrow shows the scaled variance in the EV model (\ref{eq:omega}) with $r=0.35$~fm.
    }
\label{fig:omega}
\end{figure}
Particle number fluctuations can be calculated in the grand canonical ensemble (GCE).
These fluctuations are sensitive to the interaction between particles.
The expression for the scaled variance of particle number fluctuations in the GCE EV model
is given by Eq.~(\ref{eq:omega}). The UrQMD-box calculations correspond to the  microcanonical ensemble
as the energy and charge conservation laws take place for each
microscopic state.
To study the particle number fluctuations, the conditions of the GCE should be satisfied.
These conditions can  be realized by considering a cell with the size $L_{\rm cell}$
inside the box with a size $L$. If inequalities
$1\ll \langle {N}_{\rm cell}\rangle =nL_{\rm cell}^3 \ll N=nL^3$
are satisfied,
the GCE description becomes valid for the cell with a remaining part of
the box playing a role of the thermostat.

For pointlike particles, one expects the binomial distribution to find $N_{\rm cell}$
particles inside the cell ($N_{\rm cell}=0,1,\ldots,N$). This leads to the scaled variance
\begin{equation}
\omega[N_{\rm cell}]~\equiv~\frac{\langle N_{\rm cell}^2\rangle -
\langle N_{\rm cell}\rangle^2}{\langle N_{\rm cell}\rangle}=~1-q~,
\label{eq:binom}
\end{equation}
where the parameter of binomial distribution $q=(L_{\rm cell}/L)^3$ is a probability to
find a particle inside the cell. The results of the UrQMD calculations
demonstrate in Fig.~\ref{fig:omega} a full agreement with Eq.~(\ref{eq:binom}).
Accurate estimates of particle number fluctuations require a large box with the size $L=40$~fm
and number of particles $N=3200$. Note that Eq.~(\ref{eq:binom})
gives $\omega[N]\cong 1$ at $q\ll 1$, i.e., it corresponds to the statistical mechanics
result for the pointlike particles. As seen from Fig.~\ref{fig:omega},
this is indeed the case in the UrQMD calculations. Therefore, similar
to the UrQMD pressure (\ref{box-pressure}), one does not see any signatures of the EV effects.
The EV result (\ref{PEV}) for $r=0.35$~fm is shown in Fig.~\ref{fig:omega} by the arrow.

\section{Shear viscosity of UrQMD nucleons }
\label{sec:viscosity}
\begin{figure}[h!]
\centering
\includegraphics[width=0.49\textwidth]{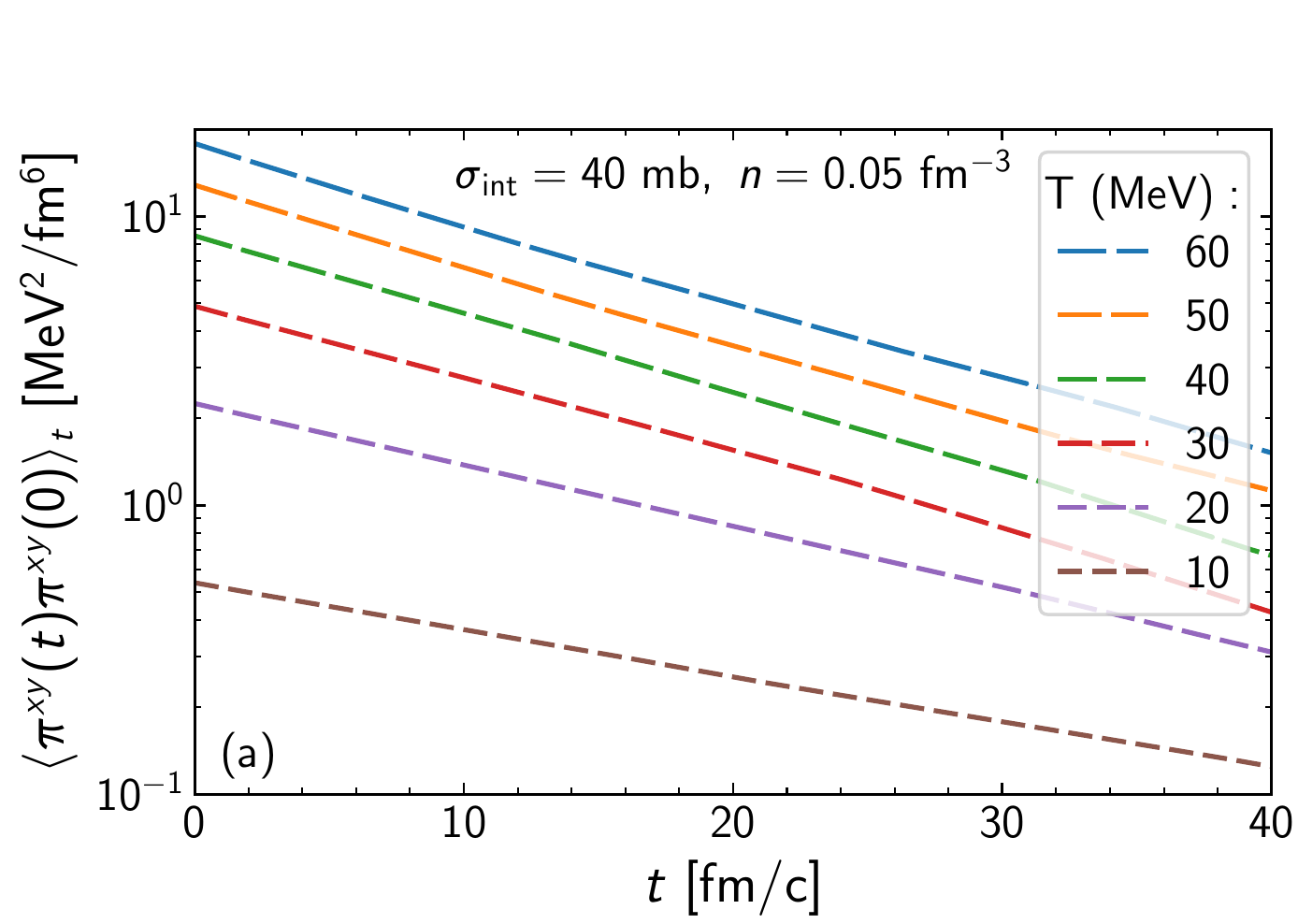}
\includegraphics[width=0.49\textwidth]{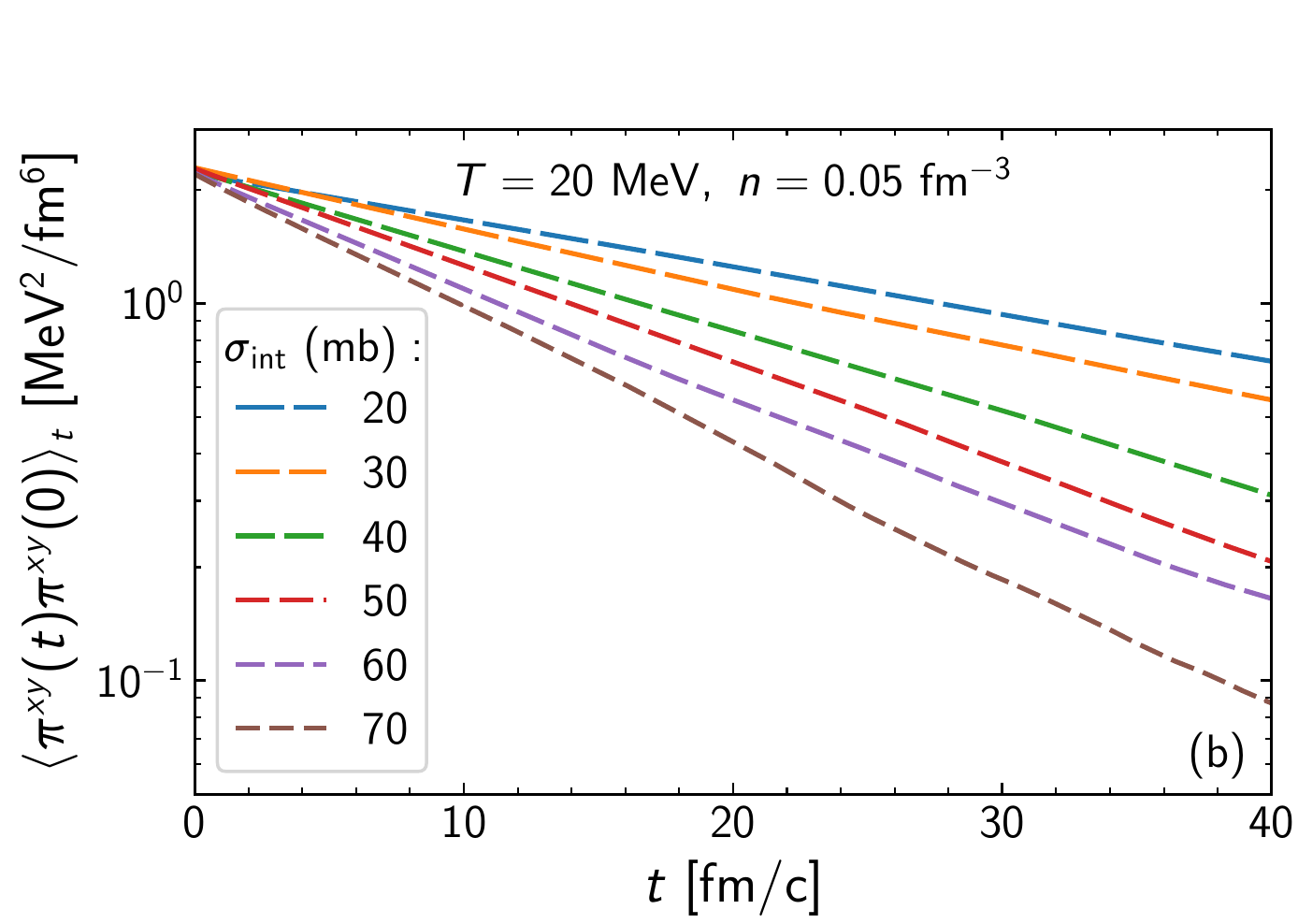}\\
\includegraphics[width=0.49\textwidth]{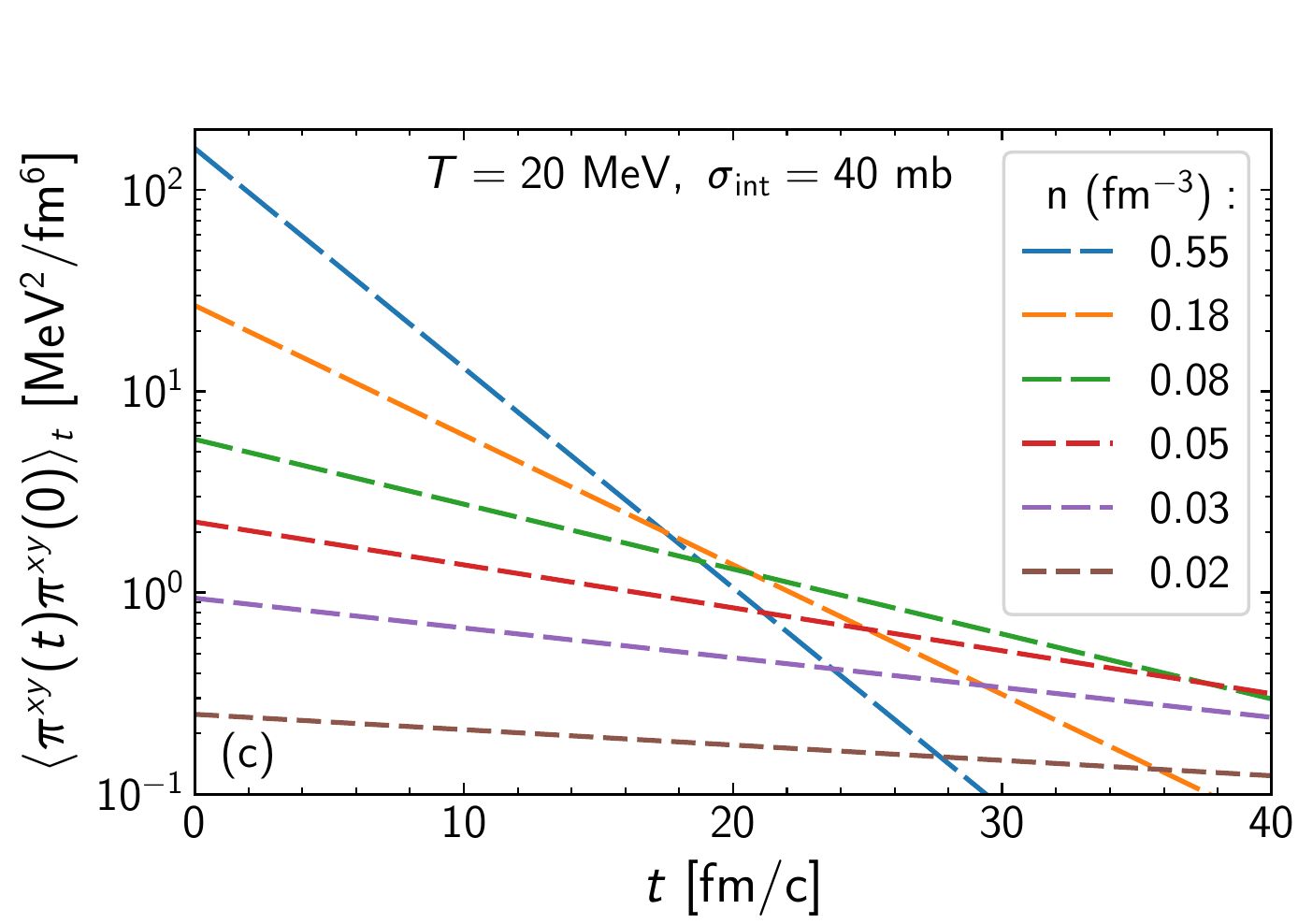}

    \caption{Self-correlators of nondiagonal components of the
    stress-energy tensor (\ref{eq:correlator}) as a function
    of time $t$ for different temperatures  (a), interaction
    cross sections  (b), and densities (c).}
    \label{fig:correlators}
\end{figure}

\begin{figure}[h!]
\centering
\includegraphics[width=0.49\textwidth]{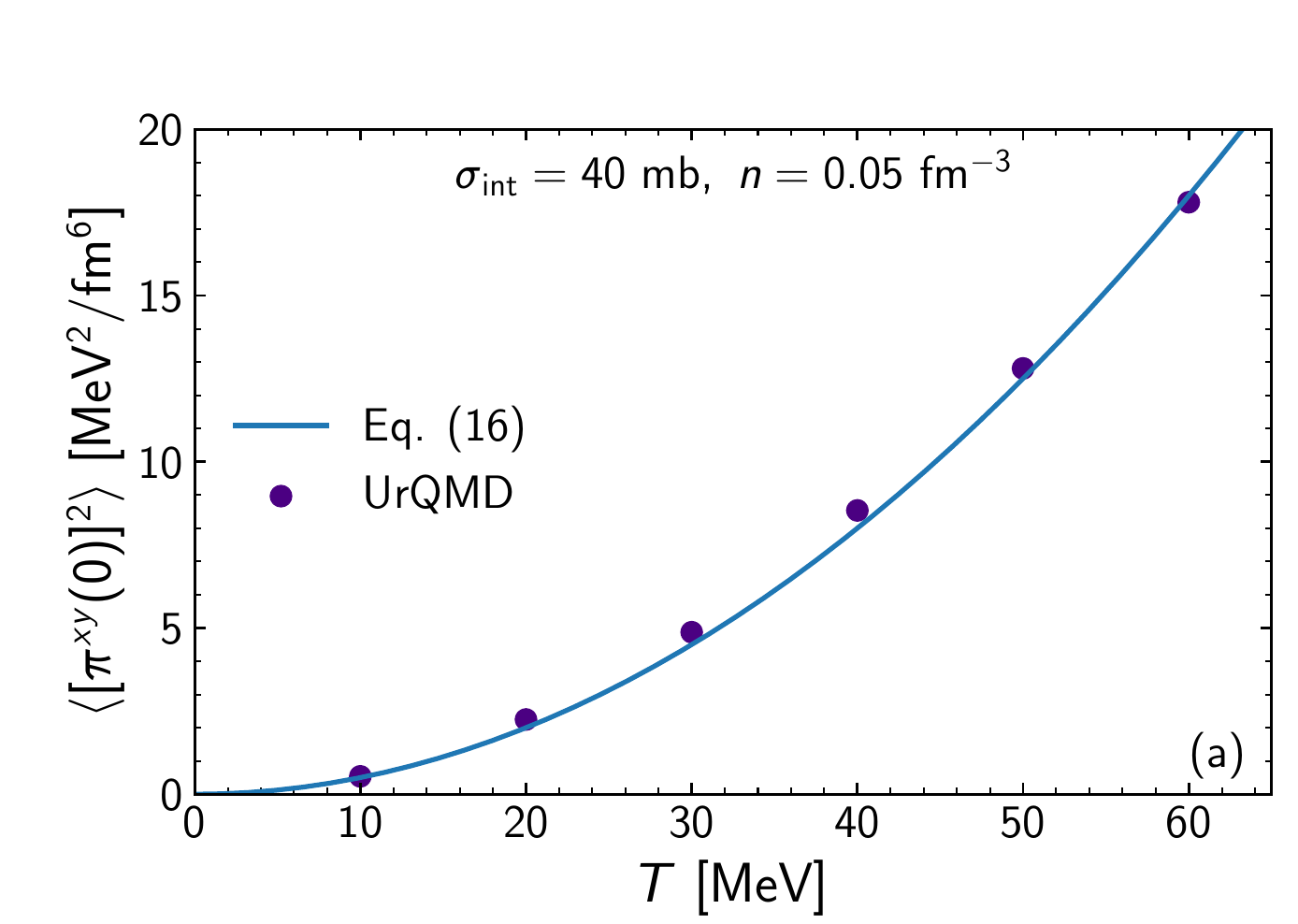}
\includegraphics[width=0.49\textwidth]{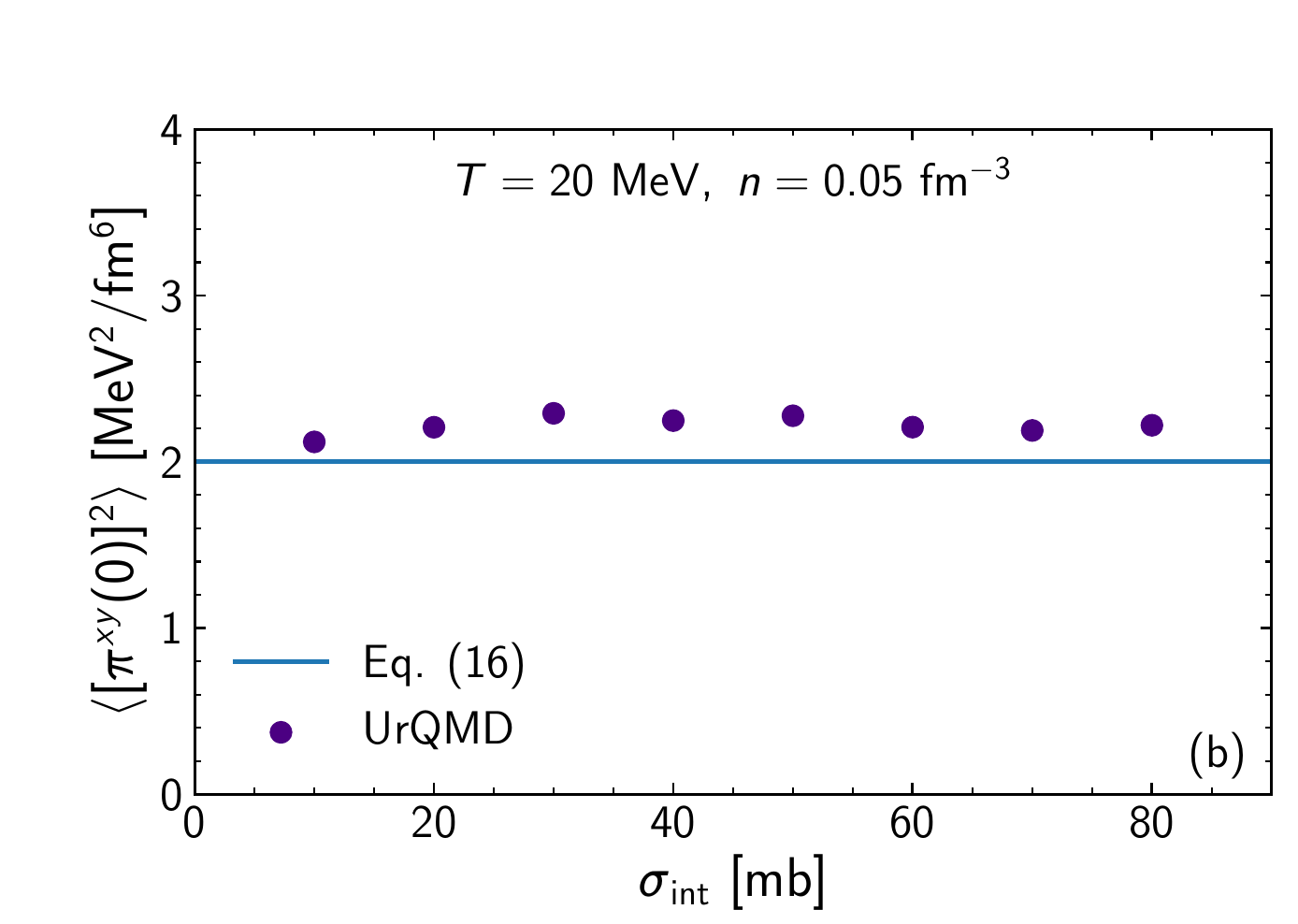}\\
\includegraphics[width=0.49\textwidth]{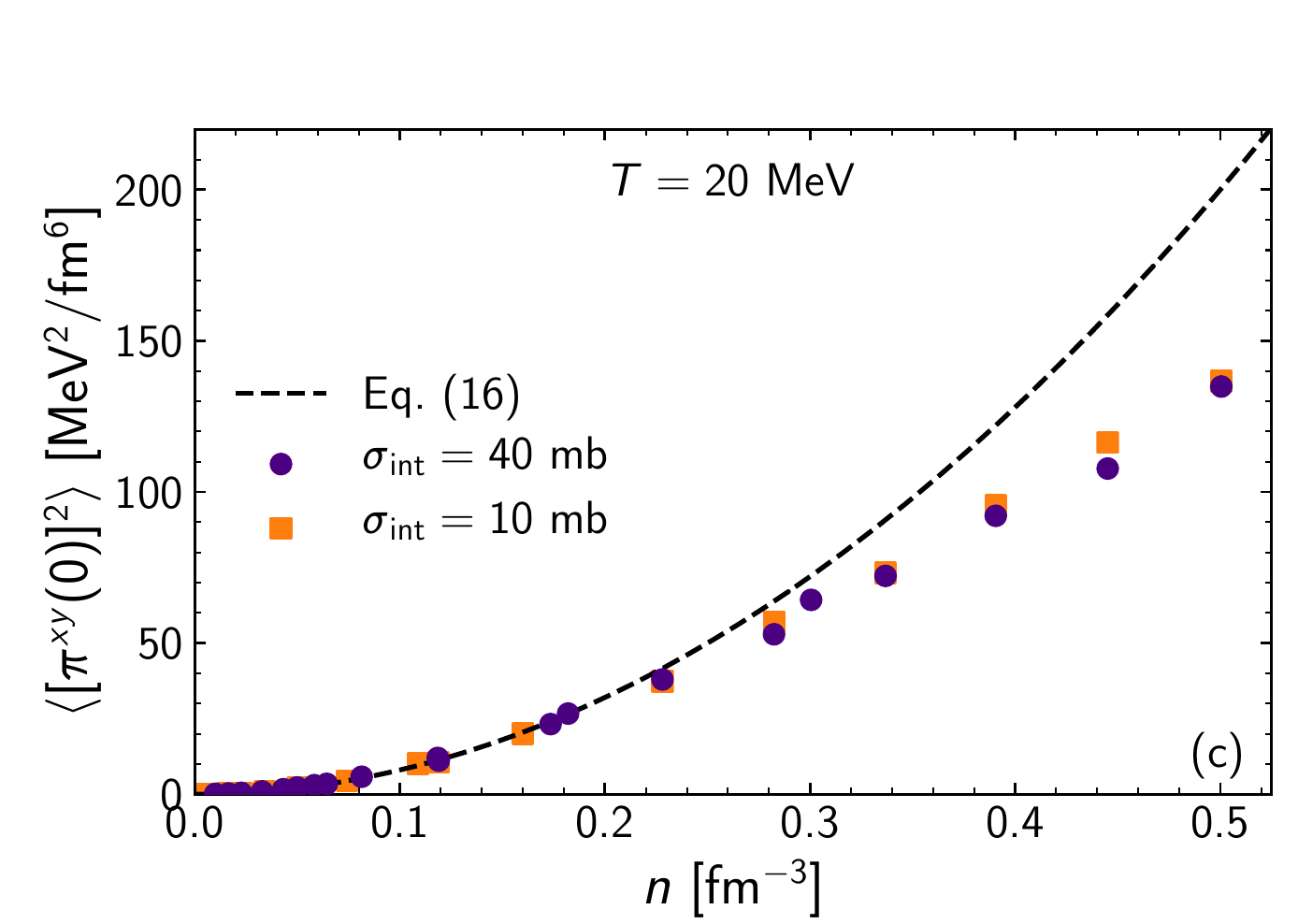}

    \caption{The amplitude of $\langle [\pi^{xy}(0)]^2\rangle$
    as a function of temperature (a),
    interaction cross section (b), and density $n$ (c).
    The lines correspond to
    Eq.~(\ref{eq:pi2}).}
    \label{fig:pi2}
\end{figure}

\begin{figure}[h!]
\centering
\includegraphics[width=0.49\textwidth]{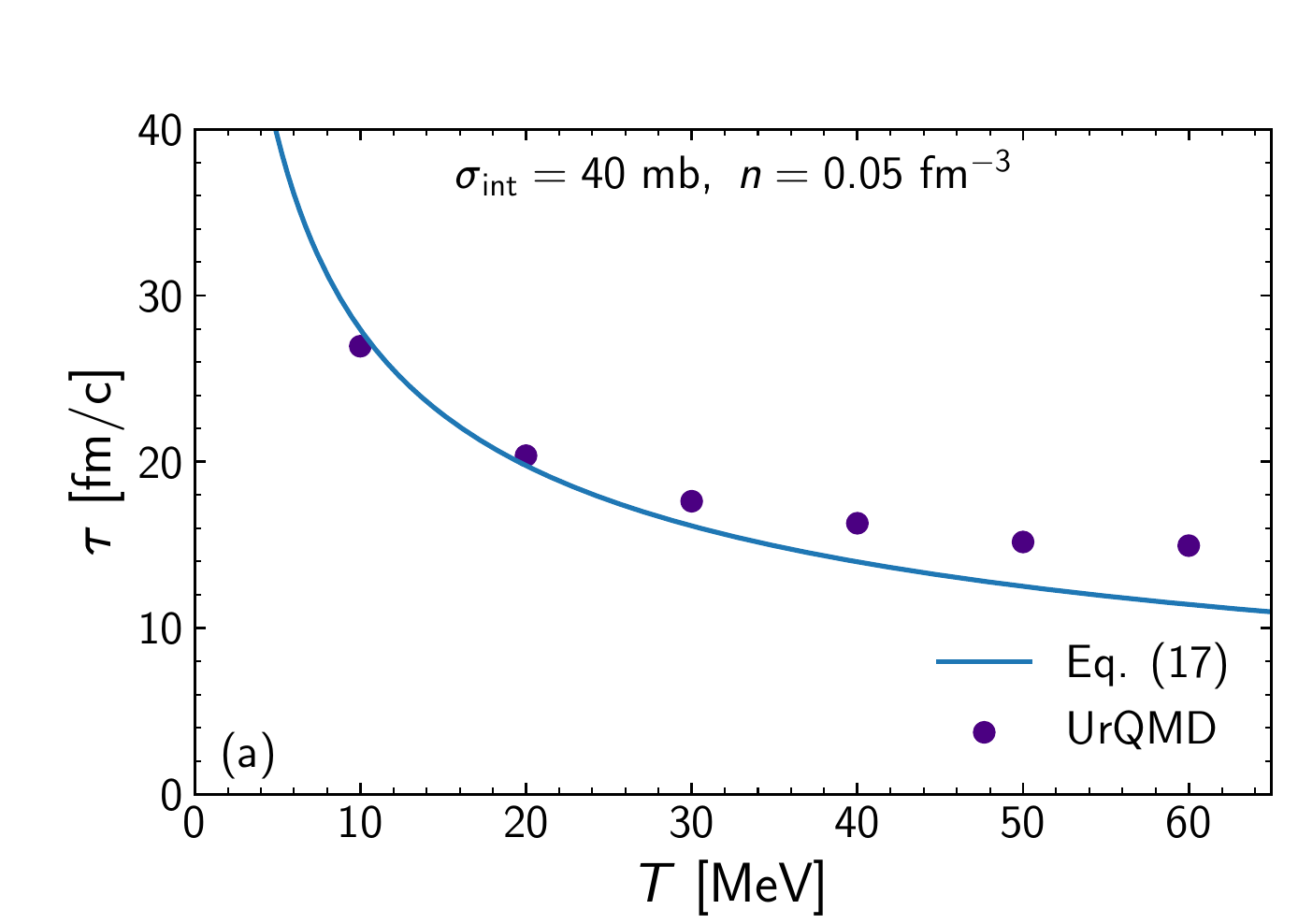}
\includegraphics[width=0.49\textwidth]{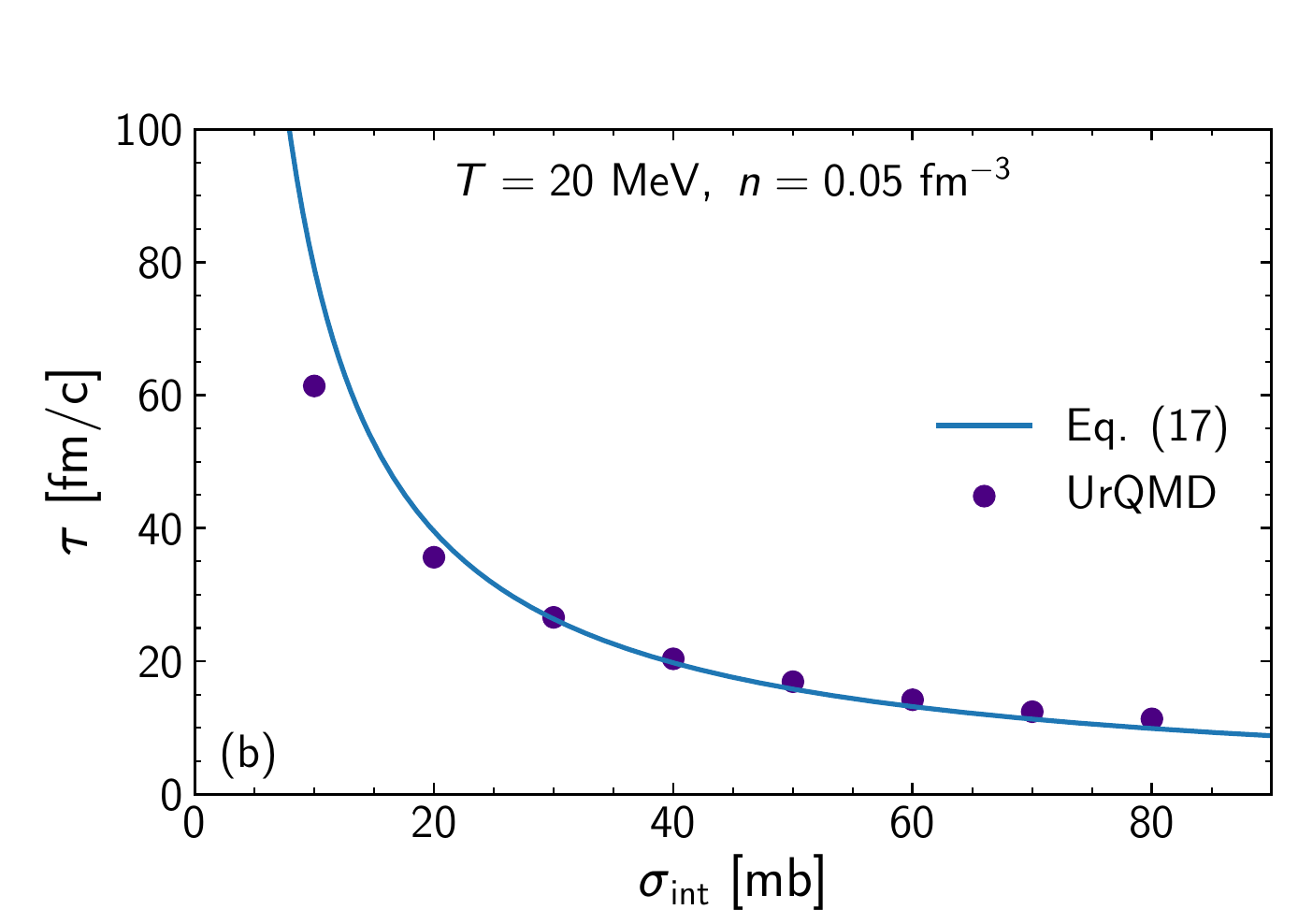}\\
\includegraphics[width=0.49\textwidth]{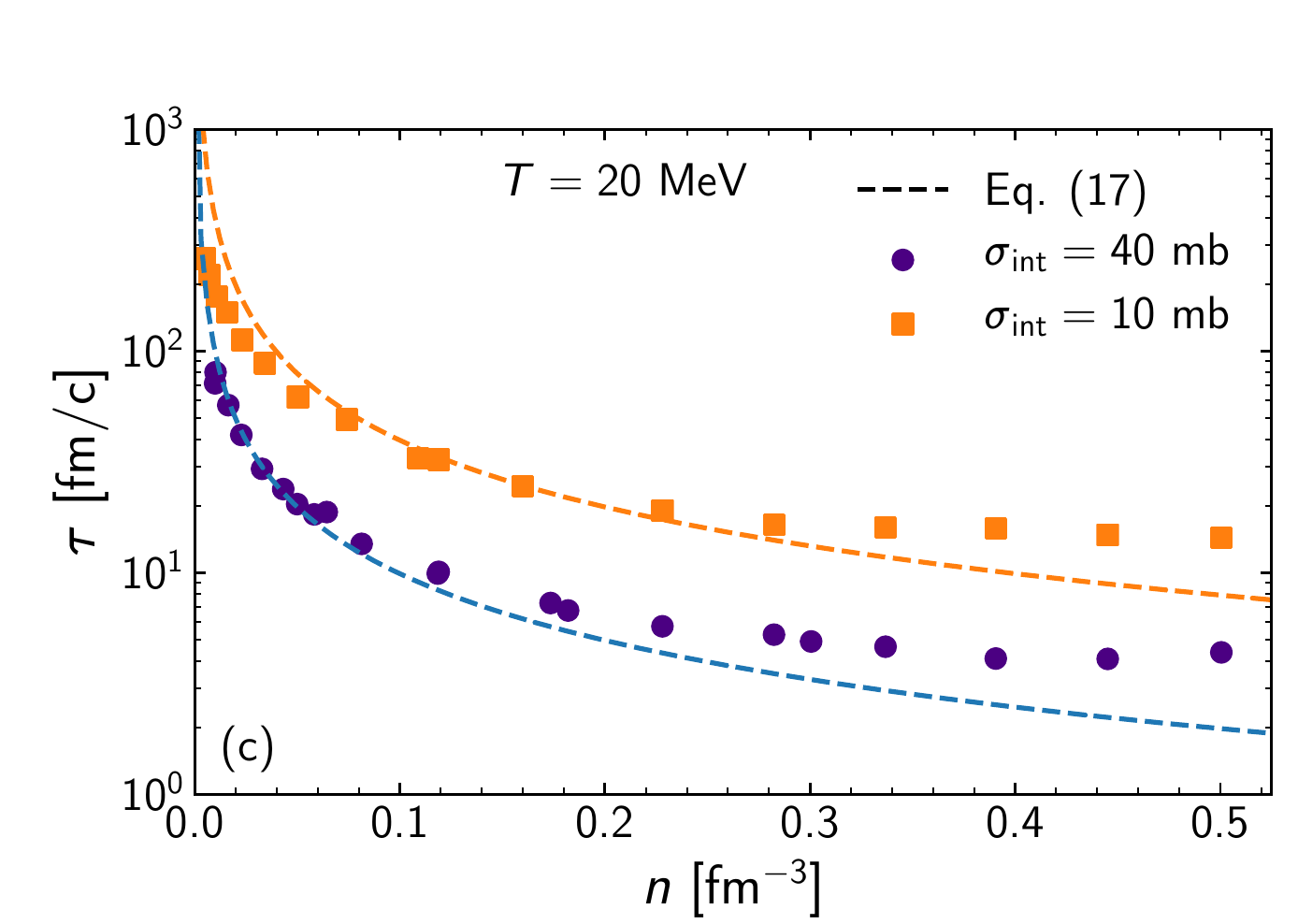}

    \caption{Relaxation times $\tau$  as a function of
    temperature, interaction cross section  (b),
    and density (c).
    The lines correspond to Eq.~(\ref{eq:tau}).}
    \label{fig:tau}
\end{figure}

To calculate the shear viscosity in the UrQMD nucleon system the
Green-Kubo \cite{Green:1954,Kubo:1957,Reichl:1980} formalism is used.
This approach has been widely applied in calculations of transport coefficients
for hadronic systems
(see, e.g.,
Refs.~\cite{Prakash:1993bt,Muronga:2003tb,Demir:2008tr,Wesp:2011yy,Ozvenchuk:2012kh,Plumari:2012ep,Wiranata:2012br,Romatschke:2014gna,Pratt:2016elw,Rose:2017bjz}).

The Green-Kubo formalism provides the following formula for the shear viscosity $\eta$:
\begin{equation}
\eta~=~\frac{V}{T}\int_0^{\infty}dt~\langle \pi^{xy}(t)\pi^{xy}(0)\rangle_t\,,
\label{eq:viscosity-GK}
\end{equation}
where $\langle \pi^{xy}(t)\pi^{xy}(0)\rangle_t$ is the time-averaged self-correlator
of nondiagonal spatial components of the stress-energy tensor $T^{ij}$,
\eq{
\langle \pi^{xy}(t)\pi^{xy}(0)\rangle_t~& =~ \lim\limits_{t_{\rm max}\rightarrow \infty}
\langle \frac{1}{t_{\rm max}} \int\limits_0^{t_{\rm max}} dt^\prime \pi^{xy}(t+t^\prime)
\pi^{xy}(t)\rangle \, ,\label{eq:correlator} \\
\pi^{ij}(t) ~ & = ~T^{ij}(t) - \delta_{ij}T^{ij}(t)\, , ~~~~~~
T^{ij}(t)  ~ =
\frac{1}{V} \sum\limits_{k=1}^{N} {p_k^i(t) v_k^j(t)}\, , \label{pij}
}
where summation is performed over all particles in a system,
$\langle \ldots \rangle$ means the ensemble averaging defined in
Eq.~(\ref{box-pressure}), and $\delta_{ij}$ is the Kronecker symbol.

Figure~\ref{fig:correlators} shows the self-correlators of
nondiagonal components of the nucleon gas stress-energy tensor calculated
in UrQMD for different system parameters. In all cases, one finds
an exponential decrease of the stress-energy correlation with time.
This correlation can be characterized by a relaxation time $\tau$ and amplitude
$\left<\left[\pi^{xy}(0)\right]^2\right>$,
\begin{equation}
\langle \pi^{xy}(t)\pi^{xy}(0)\rangle_t ~=~ \left<\left[\pi^{xy}(0)\right]^2\right> ~
\exp\left(-~\frac{t}{\tau}\right)\,.
\label{eq:corr}
\end{equation}
Such a  behavior is in agreement with a general consideration
based on the Boltzmann kinetic equation \cite{Reichl:1980}.
The initial variance of $\pi^{xy}$ presented in Fig.~\ref{fig:pi2} can be calculated analytically
by using the Maxwell-Boltzmann distribution (\ref{M}),
\begin{equation}
\left<[\pi^{xy}(0)]^2\right> ~=~ \frac{nT^2}{V}\,.
\label{eq:pi2}
\end{equation}

The relaxation time $\tau$ is expected to be
the average value of the particle propagation time
between successive collisions. It can be approximately expressed as
\begin{equation}
\tau \approx C \frac{1}{\sigma_{\rm int}\, n}~\sqrt{\frac{m}{T}}~,
\label{eq:tau}
\end{equation}
where $C$ is a factor of proportionality. From the
UrQMD calculations of self-correlators
(Fig. \ref{fig:correlators}) the value of $C$ has been extracted as $C\approx 0.58$.
Figure \ref{fig:tau} presents
a comparison of the relaxation  time $\tau$ calculated in the UrQMD box
within the Green-Kubo formalism with that given by Eq.~(\ref{eq:tau}).
From Fig.~\ref{fig:tau}~(c), one observes deviations of the UrQMD results
from Eq.~(\ref{eq:tau})
at high and small densities $n$.

From the above equations one obtains
the expression for the shear viscosity $\eta$,
\begin{equation}
\eta~\approx~ 0.58~\frac{\sqrt{mT}}{\sigma_{\rm int} }\, ,
\label{eq:eta-GK-approx}
\end{equation}
that is close to the Chapman-Enskog  result (\ref{eq:CE-sigma}), where the numerical
factor is $~5\sqrt{\pi}/16\cong 0.55$.

\begin{figure}[h!]
\centering
\includegraphics[width=0.49\textwidth]{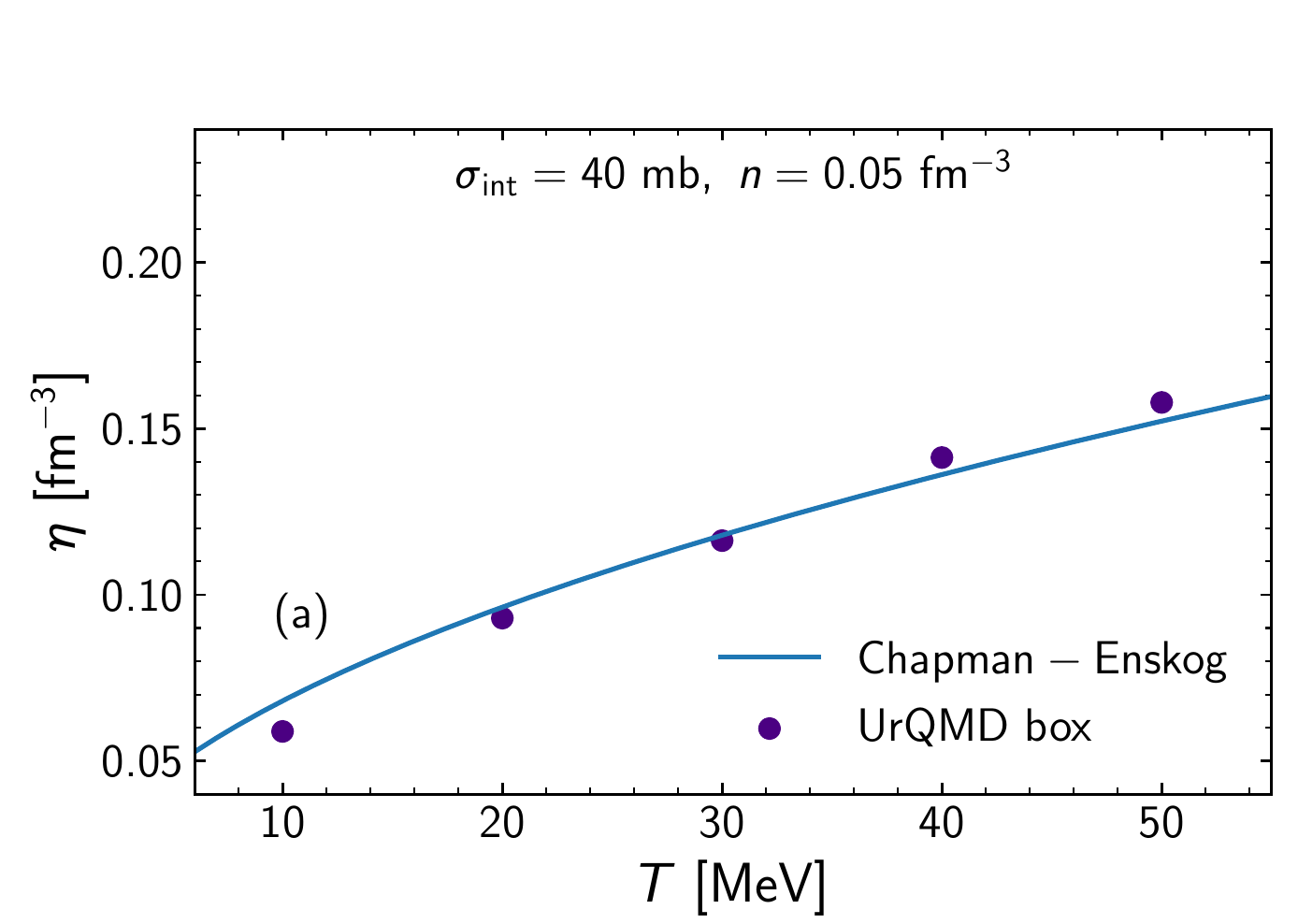}
\includegraphics[width=0.49\textwidth]{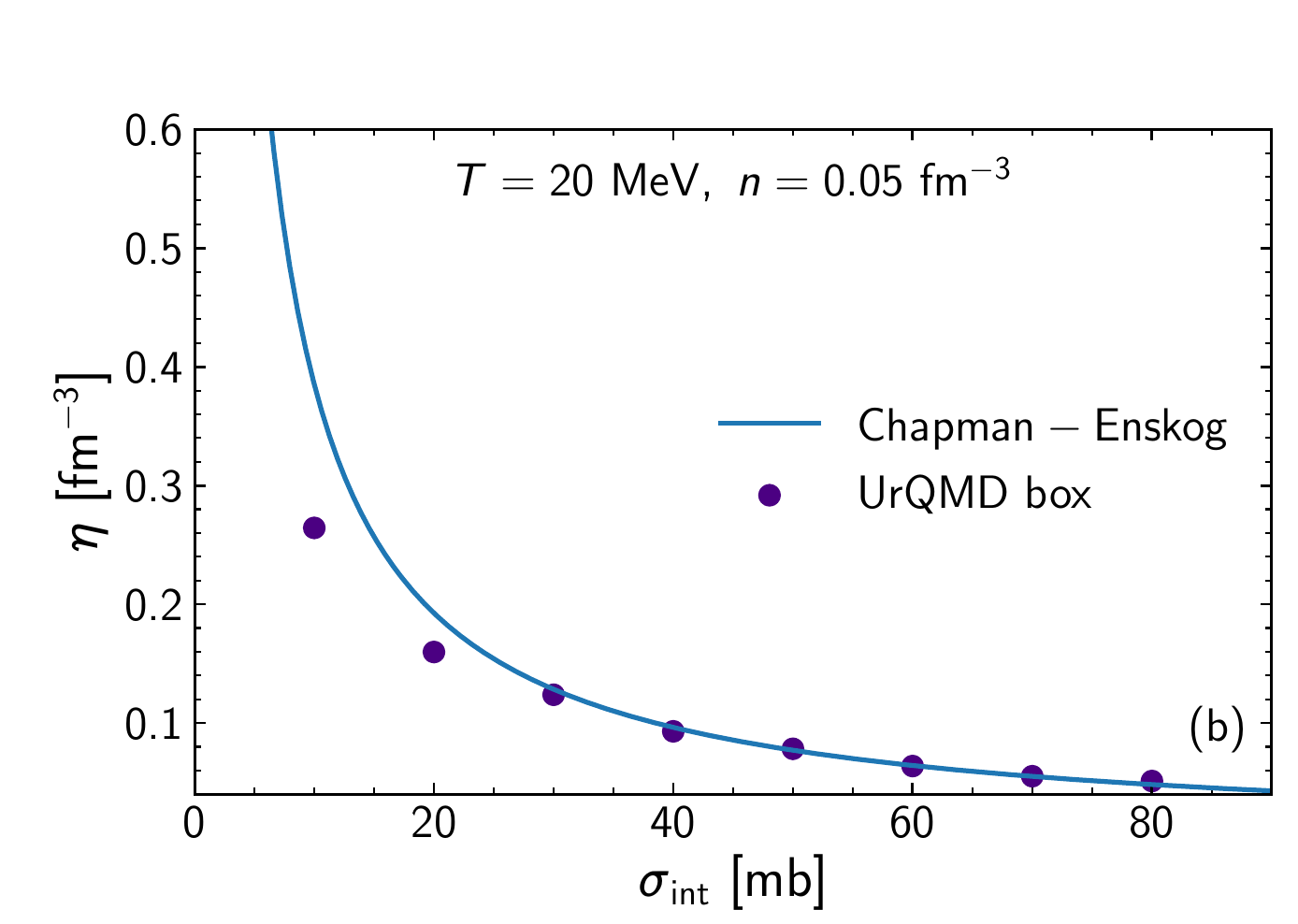}\\
\includegraphics[width=0.69\textwidth]{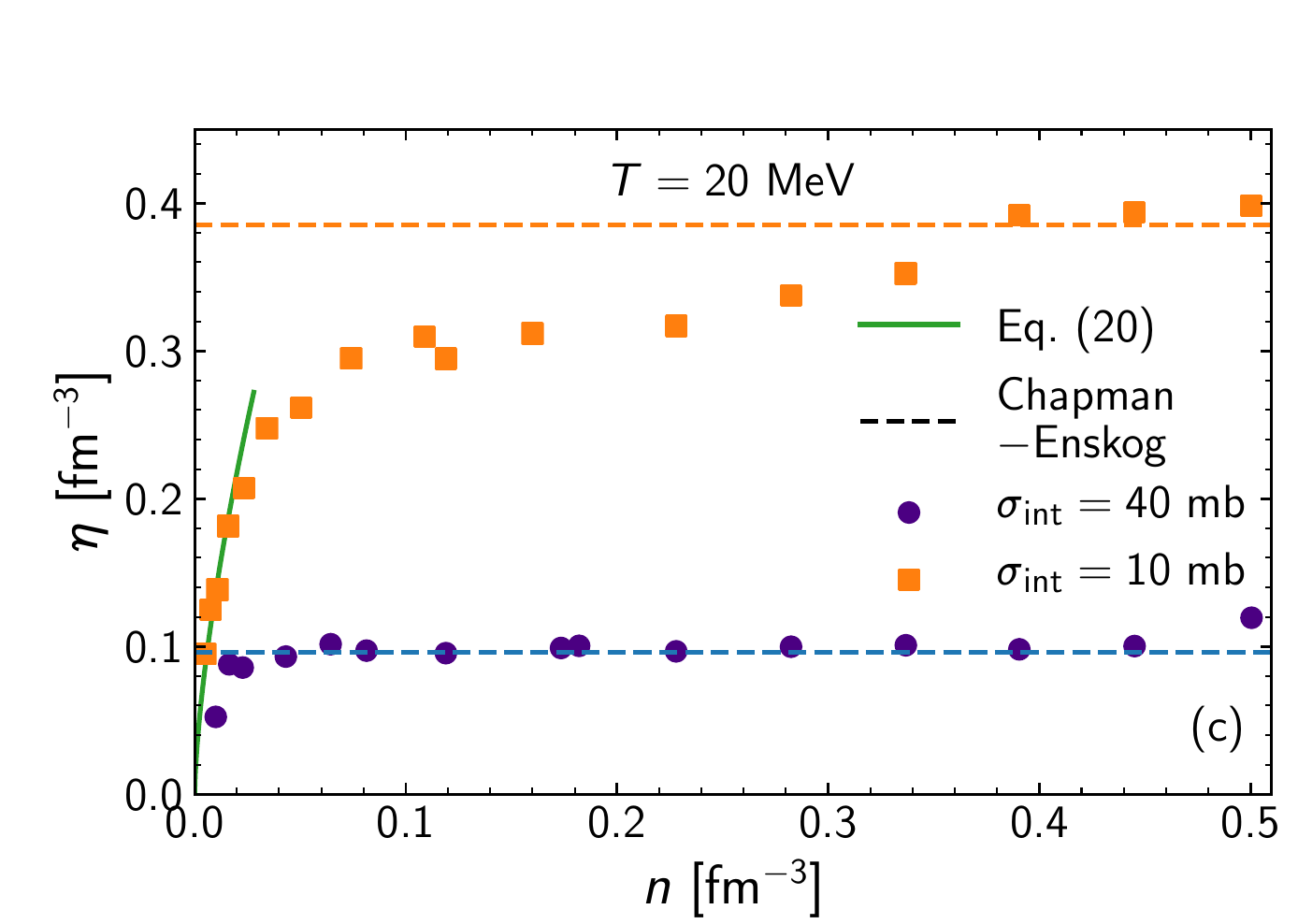}

    \caption{Shear viscosity $\eta$ as a function
    of  temperature (a), interaction cross section
     (b), and density  (c)
    obtained in the UrQMD box calculations.
    Circles and squares in (c) correspond to $\sigma_{\rm int} = 40$ mb and
    $\sigma_{\rm int} = 10$ mb, respectively.
    The Chapman-Enskog results~(\ref{eq:CE-sigma}) are shown by solid lines in (a) and (b), and dashed lines in (c).}
    \label{fig:viscosity}
\end{figure}

Figures \ref{fig:viscosity} (a) and (b) show
the UrQMD box calculations of the shear viscosity
based on the Green-Kubo formalism~(\ref{eq:viscosity-GK}). These results
are  in a good agreement with Eq.~(\ref{eq:CE-sigma}). In these UrQMD calculations,
the nucleon-nucleon elastic cross section was fixed to a certain value of $\sigma_{\rm int}$ as described in Sec. \ref{sec:UrQMD}.

As seen from Fig. \ref{fig:viscosity} (c), the deviations of $\eta$
from Chapman-Enskog results are observed at a low nucleon density.
Note that Eqs.~(\ref{eq:CE-sigma},\ref{eq:viscosity-GK},\ref{eq:eta-GK-approx})
are obtained in the FC regime which requires
$l\ll L$.  Our UrQMD calculations are performed at  fixed $N=400$,
and the nucleon density $n=N/L^3$ is changed with a variation of the box size $L$.
For the mean free path $l$, one obtains
\eq{\label{l}
l~\sim ~\frac{1}{n\,\sigma_{\rm int}}~ = ~\frac{L^3}{N\,\sigma_{\rm int}}~.
}
At (very) low nucleon density, i.e.,  (very) large $L$,  Eq.~(\ref{l}) leads to $l$ values
(much) larger than the box size $L$. This is
the RC regime.
As $l \propto 1/\sigma_{\rm int}$, the region of density for
the RC regime is wider for smaller $\sigma_{\rm int}$.
This is clearly seen from  Fig. \ref{fig:viscosity} (c).
An estimate of the shear viscosity in the RC regime
can be obtained within the molecular kinetics theory \cite{LP}.
For $l\gg L$ it results in a substitution of $l$ in Eq.~(\ref{eta}) by the box size $L$.
Therefore, one obtains
\eq{\label{eta-RC}
\eta^{}_{\rm RC}
~=~{\rm const}~ n\,m\, L\,v_{\rm th}~=~ C\,N^{1/3}\,\sqrt{mT}\,n^{2/3}\;.
}
The dependence of $\eta \propto n^{2/3}$ (\ref{eta-RC})
is shown in  Fig. \ref{fig:viscosity} (c) by the solid line.
The numerical factor $C\approx 0.58$ in (\ref{eta-RC}) appears to be the same
as in Eq.~(\ref{eq:tau}).

\section{Summary}
\label{sec:summary}
The equilibrated system of nucleons has been studied within
the UrQMD box calculations in a broad range of temperatures, densities,
and interaction cross sections. The deviations
of nucleon energy spectra from the Maxwell-Boltzmann distribution
are found at high nucleon densities. This happens because of
an implementation of the Pauli blocking in the model.
The modification of the equilibrium energy spectra does not influence, however,
on the nucleon pressure.
If the temperature parameter is defined through the average energy per nucleon,
the system pressure shows the universal ideal gas behavior, even at extreme densities.

The UrQMD nucleon interactions appear to be rather different from the elastic collisions
of hard spheres. Despite the presence of the effective nucleon size $\langle d_{\rm min}\rangle$
the UrQMD values for the system pressure and particle number fluctuations
are in agreement with the ideal gas model. Thus, no signatures
of the non-ideal gas behavior due to the excluded volume effects have been found.

On the other hand, the viscosity of the UrQMD nucleon gas
is mainly in agreement with the Chapman-Enskog results (\ref{eq:CE-sigma}) obtained for
the equilibrium system of hard spheres in the frequent collision regime. Therefore,
the details in the scattering mechanism and in the shape of nucleon energy spectra
appear to be not important: the shear viscosity $\eta$ in the UrQMD box calculations
behaves according to Eq.~(\ref{eq:eta-GK-approx}), where $\sigma_{\rm int}$ corresponds to
the elastic nucleon-nucleon cross section and $T$ is defined by the average
energy per nucleon according to
Eq.~(\ref{T}).
The difference of the UrQMD viscosity from that of
the Chapman-Enskog approach at small particle densities is observed.
It can be understood as a transition from the frequent to the rare collision
regime.

We hope that results presented in this paper will be useful for better understanding
of the equilibrium and non-equilibrium features of the transport models in their applications to A+A
collisions. They can be also helpful for further extensions and developments of the transport codes.

\section{Acknowledgements}

We are grateful to H.~Stoecker and V.~Vovchenko for encouraging comments and ideas.
Authors acknowledge discussions with D.~Oliinychenko  and J.~Steinheimer.
A.M. acknowledges the support by HGS-HIRe for FAIR, and the hospitality of the University of Oslo
where the most part of work was done. L.B. and A.M. thank the Norwegian Centre for International Cooperation in Education (SIU) for financial support, grant ``CPEA-LT-2016/10094 From Strong Interacting Matter to Dark Matter''.
LB and EZ are thankful to Norwegian Research Council for financial support,  grant ``255253/F50 - CERN Heavy Ion Theory''.
The work of M.I.G. is supported by by the Program of Fundamental Research
of the Department of Physics and Astronomy of National Academy of
Sciences of Ukraine.
A.G.M. acknowledges the support of the Program of Fundamental Research to
develop further cooperation with CERN and JINR  ``Nuclear matter in extreme conditions''
by the Department of Nuclear Physics and Energy of National Academy of Sciences of Ukraine, Grant No. CO-2-14/2017.

\end{document}